# Design and Application of Variable Temperature Environmental Capsule for Scanning Electron Microscopy in Gases and Liquids at Ambient Conditions

Ahmed S. Al-Asadi[1,2], Jie Zhang[1], Jianbo Li[1], Radislav A. Potyrailo[3] and Andrei Kolmakov[1]*

[1] The Department of Physics, Southern Illinois University at Carbondale IL, 62901
[2] Department of Physics, College of Education for Pure Science, University of Basrah, Basra, Iraq
[3] Manufacturing, Chemical & Materials Technologies, GE Global Research Center, Niskayuna, NY 12309

Correspondence should be sent to: andrei.kolmakov@nist.gov

**Abstract**

Scanning electron microscopy (SEM) of nanoscale objects in their native conditions and at different temperatures are of critical importance in revealing details of their interactions with ambient environments. Currently available environmental capsules are equipped with thin electron transparent membranes and allow imaging the samples at atmospheric pressure. However these capsules do not provide the temperature control over the sample. Here we developed and tested a thermoelectric cooling / heating setup for available environmental capsules to allow ambient pressure *in situ* SEM studies over the -15 °C to 100 °C temperature range in gaseous, liquid, and frozen environments. The design of the setup also allows correlation of the SEM with optical microscopy and spectroscopy. As a demonstration of the possibilities of the developed approach, we performed real-time *in situ* microscopy studies of water condensation on a surface of wing scales of *Morpho sulkowskyi* butterfly. We have found that the initial water nucleation takes place on the top of the scale ridges. These results confirmed earlier discovery of a polarity gradient of the ridges of *Morpho* butterflies. Our developed thermoelectric cooling / heating setup for available SEM environmental capsules promises to impact diverse needs for *in-situ* nano-characterization including materials science and catalysis, micro-instrumentation and device reliability, chemistry and biology.

**Introduction:**

Environmental Scanning Electronic Microscopy (ESEM) has manifested the great advancement in scanning electron microscopy to study the objects at elevated pressures as high as tens of hPa [1]. ESEMs are usually equipped with a commercial coolable sample holder which is connected to a thermoelectric cooling module to control the temperature inside the ESEM. The latter routinely allows to monitor the condensation and evaporation of water at vapor pressure of few hPa.[2-4] To study the samples in gases at atmospheric pressures or in fully hydrated environment in the conventional SEM instruments, environmental capsules equipped with ultrathin electron transparent membranes became recently available [5, 6] The successful application of these environmental cells was exemplified during SEM examination of fully hydrated/liquid samples such as biological tissues [5], oil emulsions [7] and functional materials [8, 9]. An additional advantage of this closed cell design is a possibility to study reactive, toxic or radioactive samples without instrument contamination. In spite of all conveniences offered by these capsules, they are not equipped with heating and/or cooling capabilities to analyze the thermal physical/chemical dynamics at nanoscale. In addition, the cell body is made out of plastic coated with thin metal layer with overall low thermal conductivity which makes it unfeasible to use commercial SEM Peltier stages to cool the entire cell.

The main objective of this communication is to report on design and tests of a cooling / heating setup for available QX-series WETSEM capsules[5, 6] to be able to observe the samples in real-time at ambient environment and yet be able to heat or cool down the sample while imaging. Butterfly wings of a *Morpho sulkowskyi* have been selected as a model object to test water condensation and evaporation process using the heating / cooling setup. The selection of these samples was motivated by the peculiar directional adhesion and super hydrophobicity of their hierarchically organized nanostructured surfaces[10, 11]. Such bio inspired materials have retained a significant interest in photonics, anti-icing and sensors [12-14]. The top surface of *Morpho* scales (Fig 1a) on micro scale has finely organized structure consisting of ordered overlapping scales[15, 16] (Figure 1b). Each scale on the other hand has a uniform periodic array of parallel ridges separated by deep grooves (Fig. 1c). At submicron scale, every ridge is made of stacked protruding lamellae Fig. 1d.  with a regular shape [17]. Overall, the butterfly wing surface has a complex 3D hierarchically  organized microsculpture with a roughness represented at nanoscopic and microscopic scales[18, 19].

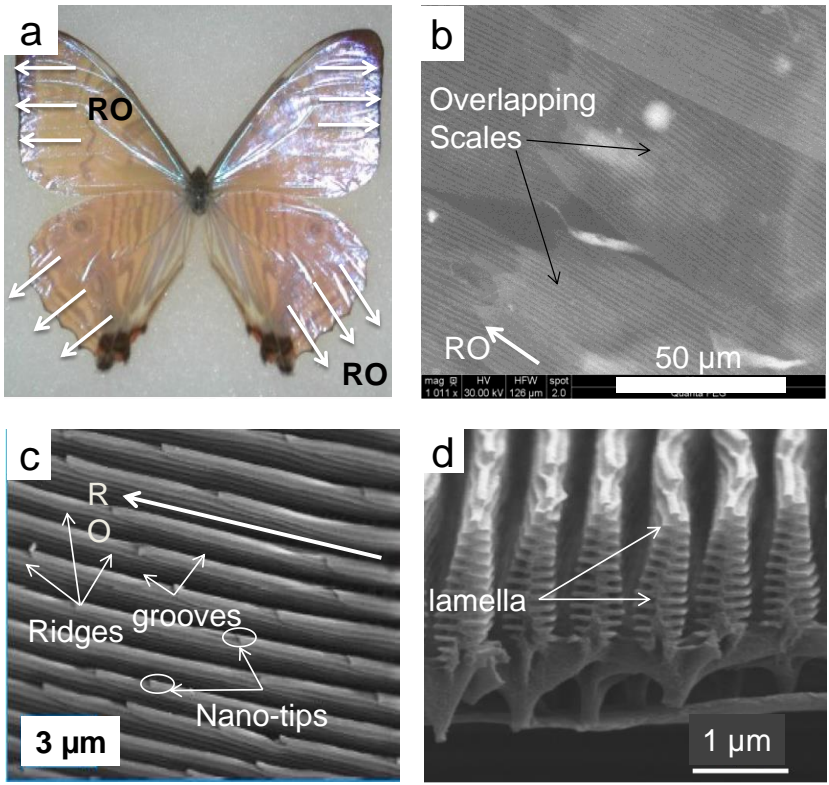

**Figure 1**:  Micro-and nanostructure of the butterfly wing. (a)  Optical image of *Morpho sulkowskyi* wing showing the Radial Outward (RO) direction. (b) SEM image of the butterfly wing surface showing the overlapping scales. (c) The microstructure of the scale with visible nano-tips, a longitudinal set of ridges and grooves observed with SEM. (d) Cross-section view of the butterfly wing surface showing multi-layer lamellae structure of the individual ridges.

In the ambient air, butterfly wings surface are known to be super hydrophobic and self-cleaning[18]. Due to inclining nature of the overlapping scales the wing, water droplets can roll off from the surface easily along Radial Outward (RO) direction [18]. Understanding of the functionalities of surfaces such as the butterfly wing could lead to a rational design of bio-inspired surfaces and coatings[18, 20] .

**Experimental:**

The standard QX-102 WETSEM capsules (Fig.2) are made of two parts: the cup, which has a sample containing compartment topped with electron transparent polyimide 145 nm thick membrane supported by a metal grid. About 15 $\mu L$ volume liquid specimen can be placed inside the compartment and imaged using 10-30 keV back scattered electrons (BSE). [8, 21, 22] The membrane can sustain more than 1 atm pressure differential and therefore the objects inside the capsule can be at atmospheric pressure. [21]

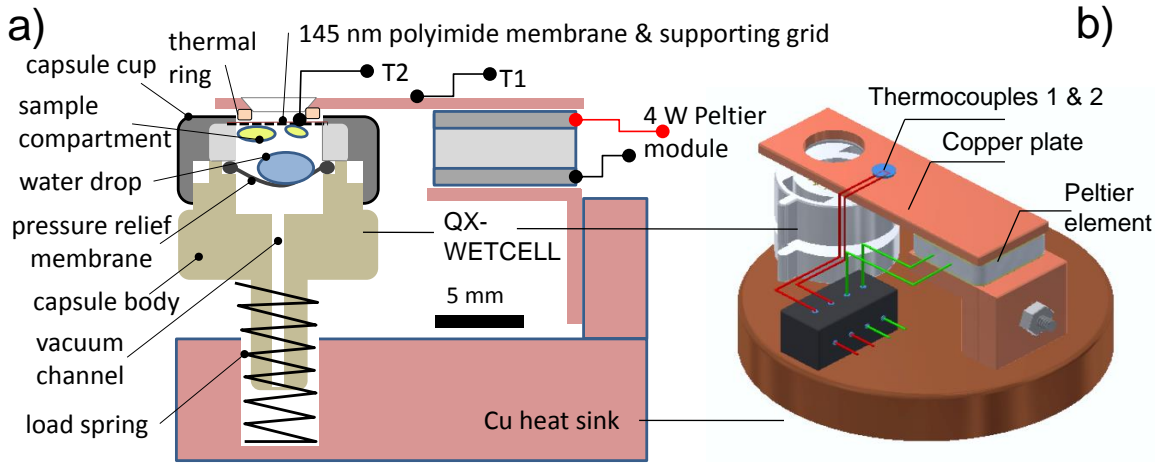

**Figure 2**: Schematic drawing of the 4.2 W thermoelectric cooling / heating setup for QuantomiX WETSEM capsules (see details in the text).

The plastic body of the QX-WETCELL is not a good thermal conductor and therefore the commercial heating /cooling SEM stages are not an optimal solution to conduct thermal studies. Instead, we decided to use the membrane supporting metal mesh on the top of the cell as a heating /cooling agent. This allowed us to deliver the heating /cooling power directly to the objects adhered to the back side of the electron transparent membrane. The overall design of the heating/cooling stage is depicted in the Figure 2. A thermoelectric $(8.4 \times 8.4 \times 2.3\ mm^3)$ element (Model 03111-9J30-20CA by *Custom Thermoelectric*) with a maximum power 4.2 W was used. The cup with membrane supporting mesh was spring loaded against cooled /heated copper plate through compressible thermal ring made of Indium (Fig 2 a). This provides good and yet removable thermal contact to the sample area. Another side of the thermal element was connected to the copper heat sink via the copper stud with a vertical slit. The top cooling plate with thermoelectric element can thus be tightly fixed at the desired height. Copper thermal sink (6 cm³) was connected to the massive microscope stage. To measure the temperature of the sample, thermocouple (K-type) has been placed on the top copper plate near to the orifice of the capsule (T1 in the Figure 2a). This thermocouple was used for calibration of the sample temperature during cooling and heating cycles. Due to limited thermal conductivity of the 150 nm thin polyimide membrane, the temperature, which is recorded by this thermocouple, does not match exactly the actual temperature of the sample inside the cell. Therefore, another thermocouple (T2 in the Figure 2a) has been temporarily connected directly to the WETCELL supporting grid and was calibrated against T1 thermocouple.

      The optical microscopy (magnification from x50 up to x1000) have been performed using Nikon L150 microscope equipped with CFI60 optics and array of imaging modes: Bright field (BF), Dark field (DF), Nomarski DIC and polarized light imaging. DF imaging appears to be best suited for *in situ* optical imaging of water condensation (evaporation) process through polyimide membrane.

SEM studies were performed on Hitachi S4500 field emission gun (FEG) microscope and environmental SEM FEI Quanta 450. The primary electron energy was selected between 5 and 30 keV. Backscattered electrons detector (BSED) was used for SEM image formation.

The small pieces (ca 1x1 mm$^2$) of *Morpho sulkowskyi* wings were cut and wet adhered to the back side of the membrane. Water droplet (ca 10 µL) was placed to the bottom pressure relief rubber membrane of the capsule in such a way that it will not touch the sample during experiment but provide nearly saturated vapor inside the compartment. The assembly can be used for imaging wet samples in optical microscope, as well as in SEM under vacuum during thermal excursions.

**Results and discussions**

*Temperature test for heating/cooling system:*

The temperature measurements were performed both in an atmospheric pressure and under vacuum by using two thermocouples to account for the thermal gradient along the setup and the heat dissipation to environment. The temperature readings in both thermocouples were measured as a function of heating/cooling (H/C) power (Figure 3). By comparing the results, it can be seen that independent of vacuum or ambient air environment, the temperature reading of two thermocouples were close within 3 $^0$C at the intermediate H/C power level (< 1W). However, further increase of the HC power resulted in temperature deviation up to 7 $^0$C degree at these two locations. The temperature on the sample is lower (higher) than on the heated (cooled) cooper plate mainly due to of the thermal resistance of an indium thermal ring.

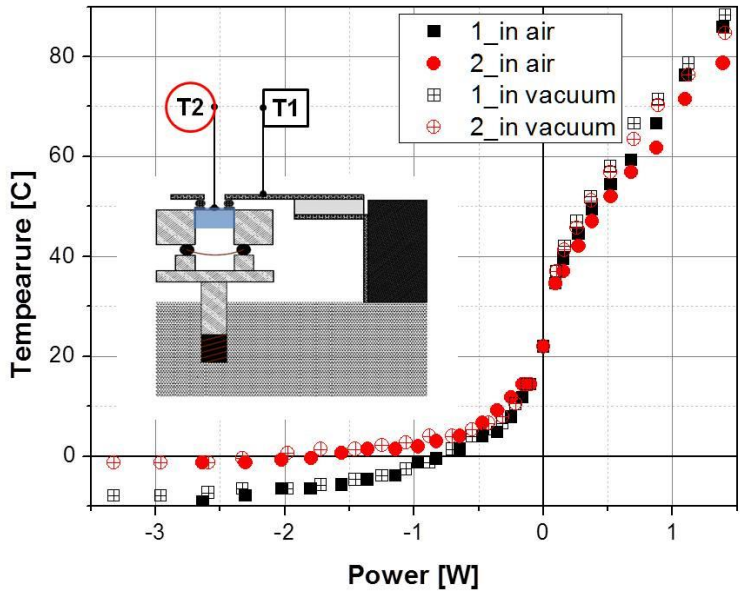

**Figure 3**: Temperature of the cell as a function of power in air and under vacuum: T1 refers to thermocouple 1 that placed on the cooper plate directly connected to the thermoelectric element. T2 represents the second thermocouple which measured the temperature on the membrane supporting mesh (closest position to the sample).

To decrease the temperature differential between the thermoelectric Peltier element and the sample, indium ring needs to be polished periodically or exchanged in order to have a better thermal contact between the copper plate and the supporting metal grid of QX capsule. Depending on the sample the maximal temperature achievable with this 4.2 W Peltier element was in access of 120 °C in vacuum and was limited by melting of In ring and polyimide membrane degradation. The minimal temperature achievable with this setup was ca -15 °C. The temperature range can be lowered further by employing higher power thermoelectric element.

**Optical microscopy tests:**

Prior to SEM studies the optical microscopy of water condensation (evaporation) process on *Morpho* scales was explored in DF mode. At ambient conditions, the temperature of the scales, water droplet inside the sealed cell was equal to the ambient temperature and the air inside the cell has the relative humidity close to 100%. Figure 4 a, e depicts dry scales with visible characteristic ridge structure. Dry scales were seen as a white objects in DF due to high scattering of the light on micro- and nanostructure features of the scale. Once the temperature of the supporting grid was decreased few degree below the dew point the massive condensation of water microdroplets on the membrane was observed (Figure 4 b-d, f). The condensation of the water on the wing proceeds first via appearance of the micron size multiple dark strips and islands elongated along the ridge direction. These islands and strips merge during continuing of condensation (Figure 4 b, c, d) making the entire wing nearly invisible in DF indicating that the space between the individual ridges become filled with water sufficiently thick to reduce the differences in refraction indexes and decrease scattering process. The comparison of high magnification images Figure 4 e and f before and after water condensation revealed that initial stages of water condensation and evaporation (not show here) proceeded strictly along ridge edges.

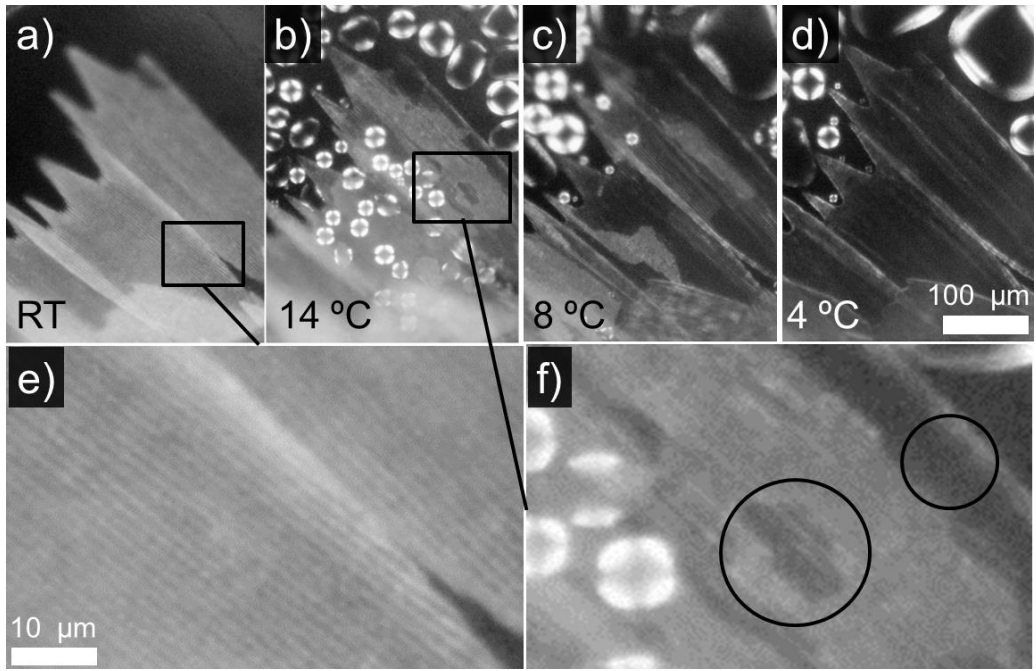

**Figure 4** Dark field optical microscopy images of the water condensation process at *Morpho sulkowskyi* wing scales recorded through 145 nm thick polyimide membrane of the sealed QuantomiX Q-102 capsule; a)-d) x50 magnification scales images of water condensation process during lowering the temperature of the wing from room temperature to 4 C. Note: the temperature reading is from T2 thermocouple which can have up to few degree of offset compared to the actual temperature of the sample; e) and f) High magnification x500 DF images of the scale before (e) and after (f) cooling of the wing to 14 C at ca 100% humidity. The referred temperatures correspond to T2 thermocouple.

In general, the butterfly wing had showed a reproducible behavior to the heating and cooling cycles in terms of the pattern formation during water condensation /evaporation. However, the temperature hysteresis was observed because of the limited ability to control the relative humidity value inside the sealed cell. To summarize, the onset of evaporation and condensation has characteristic lamella like pattern along the wing's scale what is an indication of the presence of either: (i) lamella like moieties of adsorption sites or (ii) the sites with different capacities to accumulate water or both of them. To discriminate between these two possibilities we conducted *in situ* atmospheric pressure SEM using aforementioned setup.

**Scanning Electron Microscopy temperature test**

The SEM imaging of the butterfly wing surface as a function of temperature was performed through 145 nm thick polyimide membranes using 30 eV primary electron bean and BSED detector. Similar to optical studies above, the process of water condensation on the *Morpho* scales is shown comparatively in the Figure 5 a, b and also as a movie in the supporting file. At room temperature, BSE SEM image in the Figure 5a represents a pattern of bright and dark nanostructured strips (Figure 5 a). This pattern appeared because of the variation of the back scattered electron yield from ridges (bright) and from the empty grooves (dark). Once the cooling process took place (Figure 5b), the brightness of the tops of the ridges noticeably increased, while the brightness of the bottoms of the ridges remained nearly the same (Figure 5c). We presume that the brightness increase was associated with early stages of water condensation. At this stage the water condensed predominately at very top of the ridges, what resulted in local increase of the electron back scatter coefficient and therefore increase of the brightness of the ridges (compare Figure a and b). Comparative Monte Carlo simulations of the electron trajectories[23] in the dry stack (145 nm polyimide on top of the wing substrate) versus the one with 100 nm intermediate water layer showed the increase of BSE coefficient in the latter case by ca 3-5 % what supports experimental observations. Further temperature decrease caused the thickness of the water layer on the ridges top to grow followed by coalescence of condensates between neighboring ridges forming a water bridge crossing the groove (Figure 5 b right bottom). This is an onset of complete wetting of the surface of the scale. The aforementioned condensation phenomena are in line with recently observed polarity gradient along the height of the ridge of *Morphoo* butterfly scales [24]. In particular, a biological pattern of surface functionality was found in photonic structures of butterfly scales. This pattern was a gradient of surface polarity of the ridge structures from their polar tops to their less-polar bottoms. Because of the polar nature of the very top of the ridges one can expect the preferable water condensation at the top of the ridge rather than in the bottom.

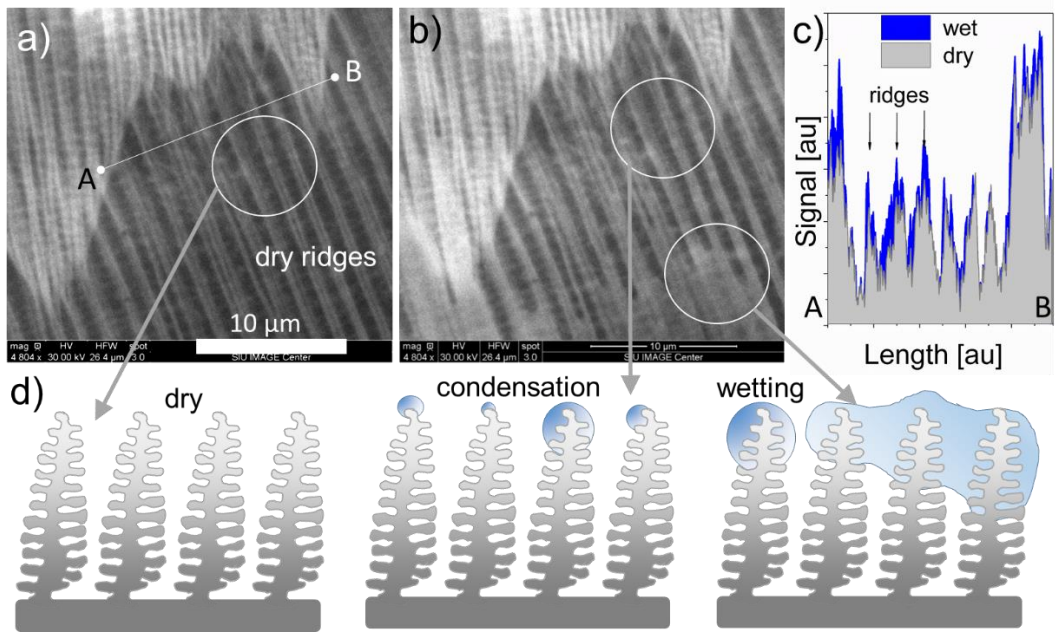

FIG 5: SEM images of the butterfly wing during cooling cycle: (a) dry scales at room temperature; (b) upon cooling to ca T=14 $^0$C, the water condensate appeared on the wing surface as brightening of the ridges. The morphology of the this thin water layer was not possible to determine due to distorting membrane effect on the resolution of the microscope; Massive condensation of water (bottom right) created thicker layer of water on the surface of the groves, what led to condensate thickening and overlap (c) The evolution of the line profile between points A and B in the Figure 5 a before and after condensation. The signal from the ridges (peaks) increased ca few percent after cooling (blue) compared to room temperature dry state (gray). The grooves gray scale values had a minor offset to evaluate this increase more precisely; (d) The proposed scenario of water condensation process correlated with observations in (a-c).

The bottom raw of the Figure 5 represents the schematics of early stages of the water condensation at ridges with polarity gradient and formation of bridged water layers covering the neighboring ridges.

It is necessary to note the existence of the beam effects on the process of water condensation and wetting on *Morphoo* scales. Zooming out of the explored area one can notice that the beam exposed areas contain less water compared to the peripheral sites of the FOV (not shown here). The latter is presumably due to heating effect of the poorly thermally conducting ridges by electron beam. This effect is commonly observed in ESEM as fast evaporation of the condensed microdroplets supported on nanostructured surfaces with reduced thermal conductivity [4, 25]. In addition, the formation of the hydrophobic areas by local beam induced carbonization cannot be excluded. The latter requires a special study.

**Conclusion:**

In this article, we have demonstrated that the simple setup can convert commercially available ambient pressure SEM capsules in to a valuable tool to study the thermally induced processes at nanoscale**.** By using thermoelectric module heating/cooling the sample area one can study *in situ* a condensation process of water on a bio-inspired complex surfaces using both optical and SEM microscopies. The obtained results supported the current hypothesis of existing polarity gradient of the ridges on the butterfly wings. Using more powerful thermoelectric elements the temperature range of the setup can be increased up to limits defined by the stability of the materials in the capsule.

**Acknowledgements**

The authors are grateful to Mr. Clay Watts for help in experiment.
**Disclaimer:**
Certain commercial equipment, instruments, or materials are identified in this paper to foster understanding. Such identification does not imply recommendation or endorsement by the National Institute of Standards and Technology, nor does it imply that the materials or equipment identified are necessarily the best available for the purpose